\documentclass[journal]{IEEEtran}
\usepackage{amsmath,amsfonts,amssymb}
\usepackage{algorithmic}
\usepackage{algorithm}
\usepackage{array}
\usepackage{subcaption}
\usepackage{textcomp}
\pdfoutput=1
\usepackage{stfloats}
\usepackage{url}
\usepackage{enumerate}
\usepackage{graphicx}
\usepackage{cite}
\usepackage{graphicx,cite,booktabs,multirow,makecell}
\usepackage{hyperref}
\hypersetup{
	colorlinks=true,
	linkcolor=blue,  
	citecolor=blue,  
	urlcolor=magenta,   
	linkbordercolor=0 0 0, 
}
\usepackage[numbers,sort&compress]{natbib}
\allowdisplaybreaks[4] 
\begin{document}

\title{Simulating Public Administration Crisis: A Novel Generative Agent-Based Simulation System to Lower Technology Barriers in Social Science Research}

\author{Bushi Xiao$^{\text{1}}$ \and Ziyuan Yin$^{\text{2}}$ \and Zixuan Shan$^{\text{3}}$ 
\thanks{$^{\text{1}}$~Department of Computer and Information Science and Engineering, University of Florida, xiaobushi@ufl.edu, United States\par $^{\text{2}}$~Department of Electrical Engineering, City University of Hong Kong, ziyuanyin3-c@my.cityu.edu.hk, Hong Kong\par $^{\text{3}}$~Department of Statistics and Actuarial Science, University of Hong Kong, u3577220@connect.hku.hk, Hong Kong}}

\markboth{Journal of \LaTeX\ Class Files,~Vol.~14, No.~8, August~2021}%
{Shell \MakeLowercase{\textit{et al.}}: A Sample Article Using IEEEtran.cls for IEEE Journals}


\maketitle

\begin{abstract}
constructing Generative Agents that emulate human cognition, memory, and decision-making frameworks, along with establishing a virtual social system capable of stable operation and an insertion mechanism for standardized public events. The project focuses on simulating a township water pollution incident, enabling the comprehensive examination of a virtual government's response to a specific public administration event. Controlled variable experiments demonstrate that the stored memory in generative agents significantly influences both individual decision-making and social networks.

The Generative Agent-Based Simulation System introduces a novel approach to social science and public administration research. Agents exhibit personalized customization, and public events are seamlessly incorporated through natural language processing. Its high flexibility and extensive social interaction render it highly applicable in social science investigations. The system effectively reduces the complexity associated with building intricate social simulations while enhancing its interpretability.
\end{abstract}


\section{Introduction}
\IEEEPARstart{S}{ocial} science research methods often simplify and abstract social phenomena to establish simplified representations \cite{2}. With the rise of computer simulations in both qualitative and quantitative social science research \cite{1,10}, computational models have become essential components of social simulation \cite{4}. Multi-agent systems, a breakthrough in computational modeling, represent distributed artificial intelligence systems consisting of interacting agents in virtual environments \cite{3,11,12}. Thus, the modeler's first task is to define the agent's perceptual and cognitive abilities.

Traditional social simulation models constructed from mathematical models such as graph theory, differential equations, and statistics are difficult to express in natural language \cite{2}, and their interpretability and practicality vary \cite{5}.

Large language models (LLMs) like ChatGPT can successfully perform a wide range of language processing tasks without pre-training, but pre-trained models can improve language understanding capabilities for more specialized tasks \cite{7}. This potential has attracted attention in computational social science \cite{6}.

The Simulacra prototype demonstrated that LLMs can be used to establish virtual communities where agents communicate in natural language \cite{8}. This opens the ``black box" of pure algorithmic agents and enables human-computer interaction through natural language.

Computational social science is a nascent field, and many theoretical explorations are still in their early stages \cite{13}. To address this gap, we propose a novel paradigm for computational social science and generative simulation: the Generative Agent-Based Simulation System (GABSS), based on the large language model GPT3.5.

What is a paradigm?

A paradigm is a general theory that provides a theoretical framework for scientific research in a specific field \cite{14}. GABSS constructs a Generative Agent program based on the human decision-making model in neuroscience \cite{16} and the human cognitive model in sociology \cite{17}. Each agent has an independent thinking cycle that includes daily planning, real-time decision-making, reflection, and experience packaging and compression for later recall.

Unlike traditional approaches that design a social network first \cite{8}, GABSS first creates independent thinking agents and then inputs each person's settings to build a social network. This paradigm is universal and creates a more general agent's thinking framework.

Social characteristics and thinking characteristics of a single agent are designed using a simplified method that can batch-modify JSON files. Once the Agent's thinking model is designed, it is theoretically embedded in the social environment \cite{9}. GABSS further simplifies the operation by reading all social environment data in natural language.

In this article, we introduce a novel generative agent-based social simulation (GABSS) paradigm to simulate public crisis events in a small town \cite{20}. We design a town based on the administrative structure of American townships \cite{18}, set up its infrastructure \cite{21,22,23}, and place more than a hundred generative agents in it. We then insert a lead pollution incident in the desalination plant at a specific timestamp, and later set a special rumor environment as a controlled experiment. We collect the natural language output of the computer simulations at each timestamp, perform statistical analysis and semantic extraction, and adopt the dual methods of quantitative and qualitative research to prove the potential of this GABSS paradigm to revolutionize computational social science.

Overall, we have achieved the following breakthroughs:
\begin{itemize}
	\item Explored research in interdisciplinary fields to design and implement a more reasonable thinking architecture of generative agents, with proven social interaction, reflection, memory, and decision-making with logical integrity.
	\item Constructed the entire agent and virtual community in natural language, increasing the model's interpretability and readability, and lowering the threshold for subsequent social science research on complex systems \cite{2}.
	\item Used text-based Json files to input different character settings and public events, enabling the simulation of any time-based event and changed the homogeneity of the traditional ABM mathematical model \cite{24,25}.
	\item Managed to mine the social and information transfer capabilities of Agent-to-Agent \cite{26}. This article explores the research value of GABSS to prove that GABSS has significant potential for social simulation and prediction of the future.	
\end{itemize}

\section{Background and Related Work}
\subsection{Agent-based Modeling in Social Simulation}
Agent-based models (ABMs) are computer simulations that represent systems as collections of digital agents \cite{11,12}. ABMs have been influential in the development of computational social science \cite{18} and can reveal and explain unpredictable social phenomena \cite{28}. For example, the Schelling model \cite{27} showed that segregation can occur even when individual agents are not highly intolerant. ABMs developed rapidly in the 1990s with the advent of large-scale computing \cite{31} and are now widely used in sociology and economics research \cite{32}.

Despite their potential for informing policy-making \cite{33, 34}, ABMs can be difficult to interpret in natural language \cite{2, 35}. Other challenges include system and simulation homogeneity \cite{36, 37}, uncertainty, and high computational cost \cite{30}. To address these challenges, researchers have proposed using agents with diverse personalities and heterogeneous system construction \cite{38, 39}, even retaining individual radical views. Work\cite{40} used agents with varying degrees of susceptibility to malicious information to simulate the spread of harmful information in social networks.

\subsection{LLM in Computational Social Science}
The development of large language models (LLMs) such as Transformer \cite{41}, XLNet \cite{42}, and GPT-3 \cite{7} has enabled powerful natural language processing capabilities like text generation, translation, and question answering. This has sparked interest in applying LLMs to computational social science (CSS) tasks \cite{13}. Recent research has evaluated LLMs on diverse CSS tasks and found they can expand CSS capabilities despite lacking professional vocabulary knowledge \cite{43}. LLMs reduce the cost of analyzing raw data, and their comprehensive databases and logical capabilities can propose new models and hypotheses \cite{6}.

A key innovative direction is LLMs' simulation capabilities \cite{46}. LLMs can generate synthetic interview responses that position real interviews in advance \cite{45}, though cannot replace real survey data. LLMs also generate synthetic social media content that is difficult for humans to distinguish from real content \cite{44}. Overall, LLMs have many promising applications in CSS, though fine-tuning may be needed for optimal performance on specialized tasks. More research is needed to responsibly leverage LLMs' strengths while mitigating their limitations.

\subsection{Generative Agents}
The concept of `social simulacra' using large language models (LLMs) to generate human-like conversational agents was proposed by Park et al. (2022) as a way to design computational social systems \cite{8}. Subsequent work has built on this to develop 'Generative Agents' (GAs) that can simulate realistic social behavior \cite{9,48,52}. GAs are trained on large social media corpora to enable natural language communication. Human evaluations show promise for using GAs in education, entertainment, and social science.

Recent advances like chain-of-thought prompting have improved LLM reasoning abilities \cite{47}, enabling applications like GABM which embeds GAs into epidemic models \cite{51}. However, concerns remain about potential biases, reproducibility, interpretability, and ethics of both LLMs and GAs \cite{49,50}. Limitations include memory degradation and excessive rigidity when fine-tuned \cite{9,52}.

Overall, GAs and LLMs are powerful but imperfect tools. More research is needed to responsibly develop and evaluate their capabilities for simulating human behavior and expanding computational social science methods. But they may help address limitations of traditional agent-based modeling like agent homogeneity \cite{13,14}.

\section{Generative Agent-Based Simulation System}
Our Generative Agent architecture builds on traditional ABM methods which start with simple agent models then increase complexity \cite{2}. We establish an independent cognitive system for each agent, simulating human memory and decision-making processes \cite{16,53,26}. This proposer-predictor-actor-critic framework guided by large language models \cite{15} aims to precisely model human behavioral selection mechanisms.

The core architecture contains interconnected planning, decision, reflection, and memory systems that exchange natural language data. This framework leverages large language models' semantic capabilities \cite{15} to mimic human cognition. Each agent's thinking system is independent, so social interactions uniquely transmit agent-to-agent information \cite{53,26}.
\begin{figure}[htpb]
	\centering
	\includegraphics[width=0.9\linewidth]{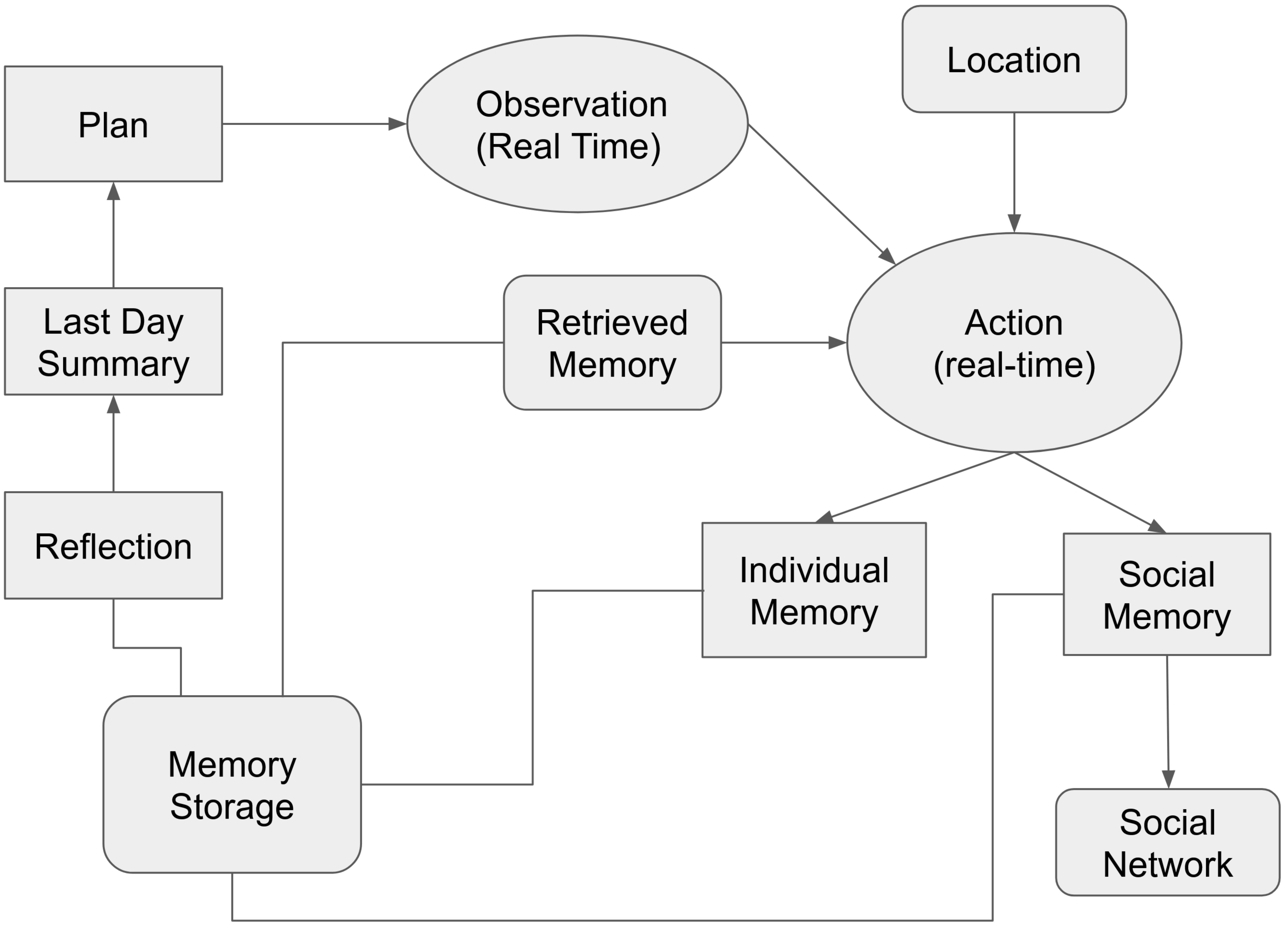}
	\caption{Architecture of Single Generative Agent.}
	\label{fig:fig1}
\end{figure}

Overall, the thinking framework uses natural language processing to increase the cognitive fidelity of individual agents compared to traditional ABM methods. The independent yet interconnected agent systems aim to realistically simulate how humans think and behave in social contexts.

\subsection{Planning-Decision System}
The goal-driven yet reactive decision process\cite{16} aims to mimic human prioritization and cost-benefit analysis. The planning system generates long-term goals, but the decision-making system determines the agent's ultimate actions via continuous deliberation adjusted for real-time conditions. 

The planning system will refer to the read-in agent character settings, combine each day's reflections with the memory data in the memory system, and generate the next day's plan at zero o'clock every day: the agent will generate an action for every hour of the next day.

The generative agent's decision-making system is central to simulating intelligent behavior and runs in a loop every 5 minutes. It references the overall plan from the planning system but crucially makes real-time decisions based on the environment. The key decision is choosing the agent's next location.
\begin{figure}[htpb]
	\centering
	\includegraphics[width=0.99\linewidth]{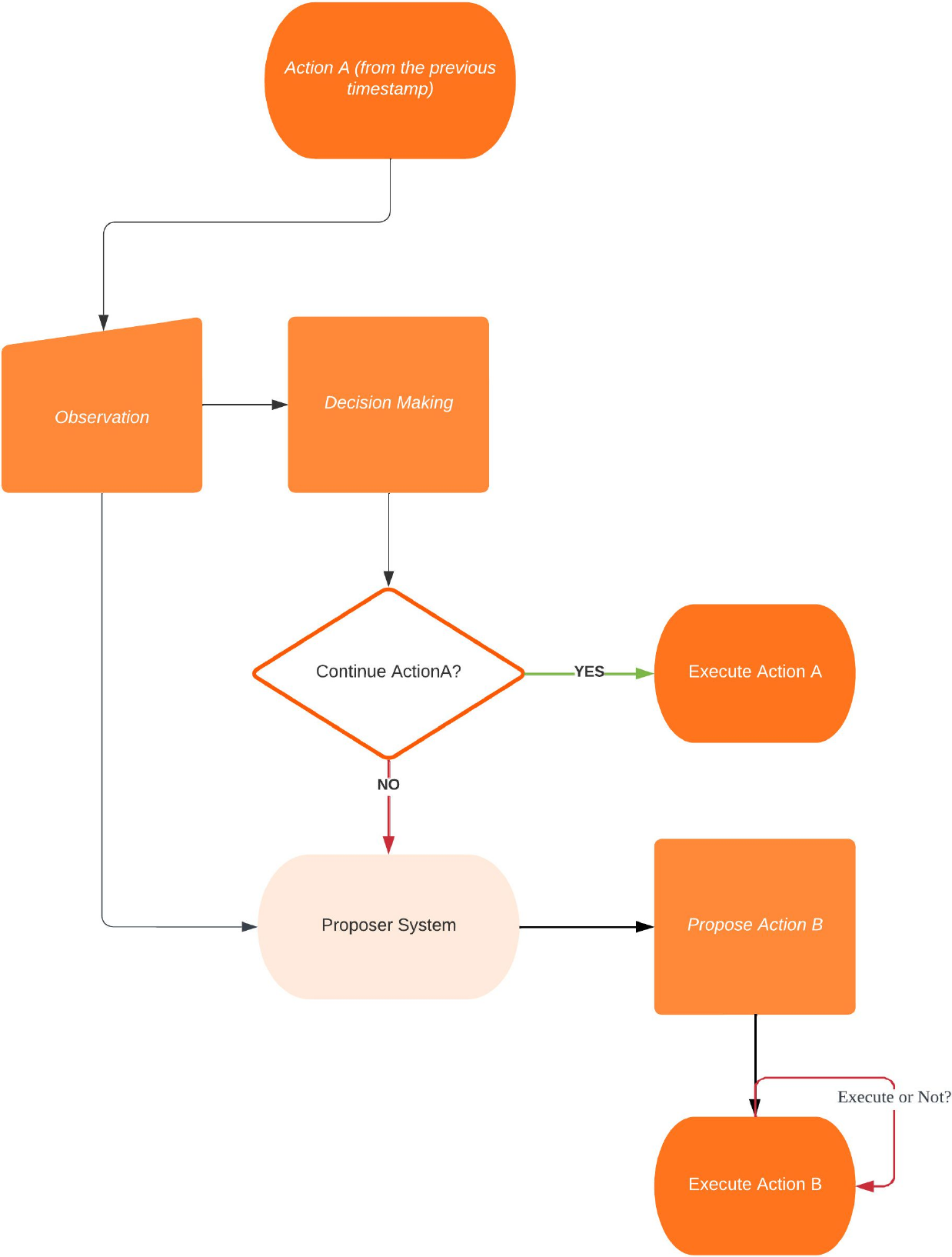}
	\caption{Decision Making Structure.}
	\label{fig:fig2}
\end{figure}

The system evaluates the necessity and distance cost of each potential target node, prioritizing those in the same town-area to minimize time costs. It perceives the environment, reasons about options, and decides on an action, following the perception-cognition-decision loop of human cognition. Through the ``observation" mechanism, the agent will traverse the text descriptions of all interactive items in the leaf node, predict its interactive value, and decide whether to interact with it.

The decision-making system has two execution options: ``execute" or "don't execute" a proposed plan \cite{2}. If declined, it proposes alternatives until acceptance. Due to programming constraints, only one decision can be made per timestamp.

During execution, the system continuously observes the environment and may interrupt ongoing behaviors for more important ones. This models how human intuition and impulses can shift to rational plans with deeper deliberation.

\subsection{Information Transmission in Social Interaction}
Social information transmission greatly influences human social cognition \cite{54,55,56}. Our GA architecture incorporates two key social aspects:
\begin{enumerate}[1.]
	\item A social memory network based on personas and memories that stores known relationships.
	\item Observation and evaluation of new agents to establish connections. Agents at the same location assess if social interaction is worthwhile.	
\end{enumerate}

A ``request-response" mechanism governs interactions:

Agent A evaluates if Agent B merits communication. If so, A requests interaction. B evaluates A and either accepts, starting a dialogue, or rejects, ending the attempt.

Key information is extracted into social memories. By combining preset and dynamic relationship building, this system aims to mimic the expandability of human social ties.
\begin{flushleft}
\textit{Social memory update after Agent ``Karl" chatted with his grandson Agent ``Tuvva":}

\textit{I know from Tuvva that: Tuvva learned that connecting with Grandpa Karl can lead to new and exciting experiences., Created: Day 0 08:00, Last visited: Day 0 08:00}
\end{flushleft}

\subsection{Temporally-Aware Memory Retrieval}
In enhancing the capabilities of generative agents, the facet of memory retrieval holds a quintessential position, offering a contextual lens to perceive historical interactions and experiences. The methodology delineated herein endeavors to enrich the memory retrieval schema by intertwining temporal information with memory vectors. This synthesis aspires to orchestrate a more discerning retrieval predicated on both semantic and temporal closeness.

\subsubsection{Memory Vector Generation}
The journey commences with the translation of each memory, narrated in natural language, into a vector representation via a pre-trained BERT model. Accompanying each memory is a timestamp marking the occurrence of the event. For a memory with description ($d$) and timestamp ($t$), the vector representation ($v$) is garnered as follows:
\begin{equation}\label{1}
v\ =\ BERT(d)
\end{equation}

\subsubsection{Temporal Encoding}
\begin{figure*}[htpb]
	\centering
	\includegraphics[width=0.98\linewidth]{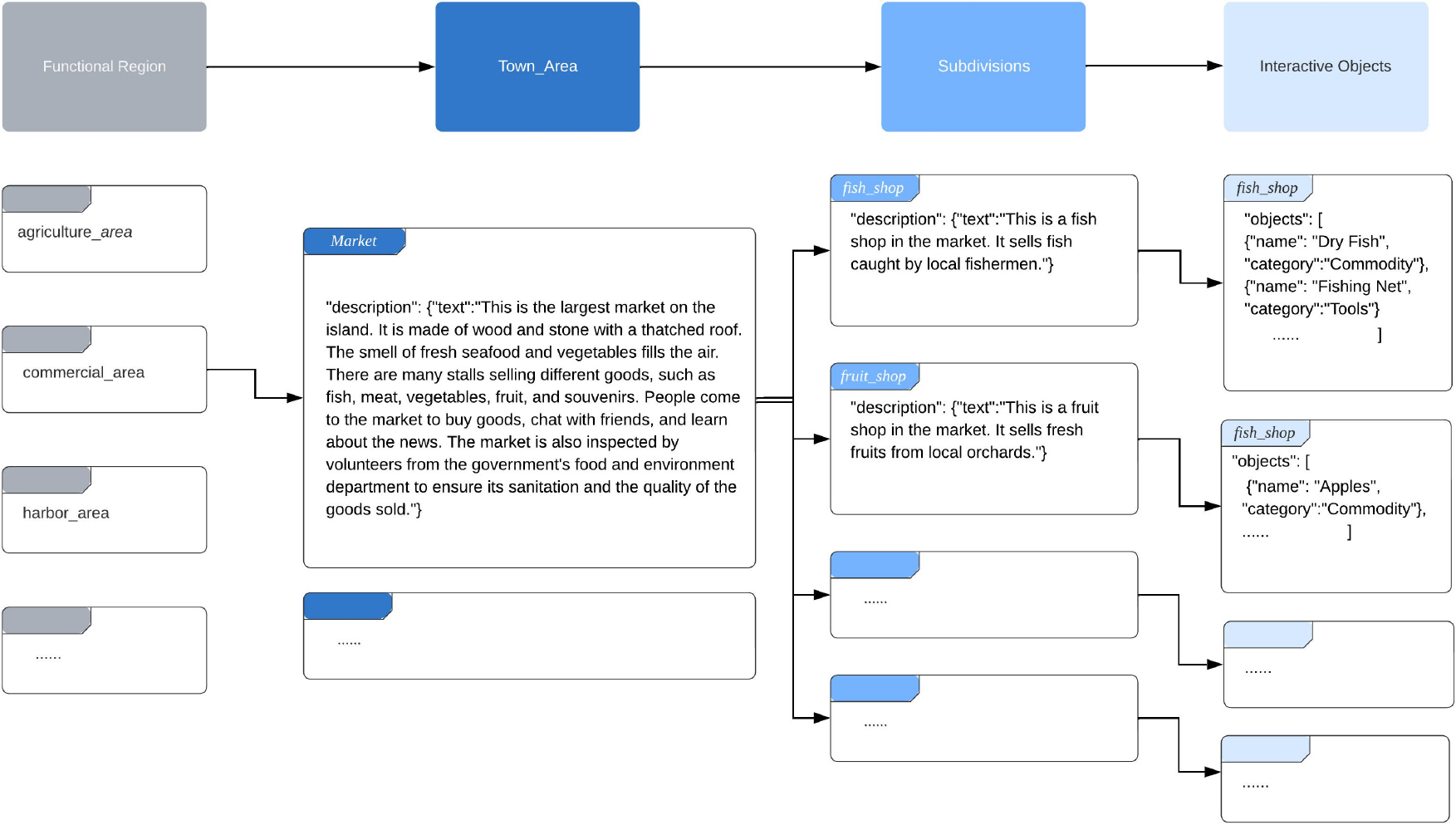}
	\caption{Hierarchical Structure of Social Environment.}
	\label{fig:fig3}
\end{figure*}
Temporal essence is encapsulated by metamorphosing the timestamp $t$ into a numerical avatar represented in seconds, denoted as $G$. A hyperparameter $\alpha$ is ushered in to modulate the time's sway over the vector representation. The encoded temporal datum $T_{encoded}$ is sculpted as:
\begin{equation}\label{2}
\ T_{encoded}\ =\ G\ \times\ \alpha\ 
\end{equation}

\subsubsection{Temporally-Aware Memory Vector}
The temporally-aware memory vector $V_{time-awar}$ burgeons by the multiplication of $T_{encoded}$ with the vector representation $v$ of the memory:
\begin{equation}\label{3}
V_{time-awar}\ =\ v\ \times\ T_{encoded}
\end{equation}

This choreography ensures a closer affinity in the vector space for memories birthed in temporal proximity.

\subsubsection{Importance Scoring}
An additional layer of refinement in the retrieval odyssey is introduced via an importance score attributed to each memory. This score, journeying from 1 to 10, bifurcates mundane events from those bearing significant import. The importance score unfurls directly from the language model output as an integer value, nurtured by the description of the memory.

\subsubsection{Query Vector Preparation}
The query vector $q$ is cultivated through a similar alchemy, where the description is transmuted into a vector via BERT, then mulled with the temporal encoding of the query time, akin to the memory vector generation.
\begin{equation}\label{4}
q\ =\ BERT(query_{description})\ \times\ (query_{\times\alpha})
\end{equation}

\subsubsection{Memory Retrieval}
The act of memory retrieval is orchestrated by computing the Euclidean distance between the query vector $q$ and each temporally-aware memory vector. The similarity score emerges as the negative of the Euclidean distance:
\begin{equation}\label{5}
Similarity\ Score\ =\ ||q\ -\ V_{time-awar}||
\end{equation}

With the ambition to intertwine the importance score in the retrieval narrative, another hyperparameter $\beta$ is unveiled. The final score for each memory is composed as:
\begin{equation}\label{6}
Final\ Score=Similarity\ Score+(Importance\ Score\times\beta)
\end{equation}

The memories are then arrayed based on the final score, with the pinnacle-ranked memories being retrieved as the most pertinent memories for the tendered query.

This methodology unfurls a robust scaffold for temporally-aware memory retrieval, empowering the generative agent to delve into relevant memories with a nuanced understanding of the temporal and semantic milieu, markedly amplifying the agent's performance and interaction finesse. Through the meticulous melding of temporal information and importance scoring, this methodology heralds a stride towards more sophisticated memory retrieval machinations in generative agents, contributing significantly to the burgeoning narrative of human-like artificial intelligence systems.

\section{Virtual Social Environment Embedding}
In this section, we elaborate on the construction of the entire virtual society by referring to the social and population structures of real-world towns. The simulated town A is located on an island, allowing us to reduce the impact of in-migration and out-migration in short-term simulation \cite{57}. The agent settings also reference the demographic structure and characteristics of similar societies \cite{58}. As a township-level administrative district, we refer to the administrative structures of American towns \cite{18,60}.

\subsection{Township model}
Referring to general township structure\cite{23, 42}, town A is divided into multiple areas (town\_area) by function. Each town\_area represents a functional area like commercial or residential. Different town\_areas connect through preset roads.

Subdivisions: Each town\_area subdivides into multiple subdivisions, representing components like a commercial street block. Each subdivision has a natural language description to facilitate agent understanding of functional attributes.

Interactive objects: Within subdivisions, interactive objects categorize into types with attributes. Agents observe objects in the subdivision and choose to interact.

Through multi-level regional design, interactive objects, and agent autonomy, we built a fully-functional, near real-world virtual town society. This provides scenario support for agent-based public administration simulation.

\subsection{Demographic Structure of Agents}
The population configuration of this virtual society references the demographic ratios of archetypal island towns \cite{57}: young adults aged 18-35 comprise 34.9\%, middle-aged residents aged 35-60 constitute 43.1\%, and the elderly populace is relatively small. Considering most towns with low development levels concentrate in primary industries, employment focuses on agriculture and animal husbandry (24.3\%), fisheries (12.1\%), services (22.4\%), and light industry (13.1\%). A limited number of unemployed individuals (6) reflect unstable factors.
\begin{figure}[htpb]
	\centering
	\includegraphics[width=0.92\linewidth]{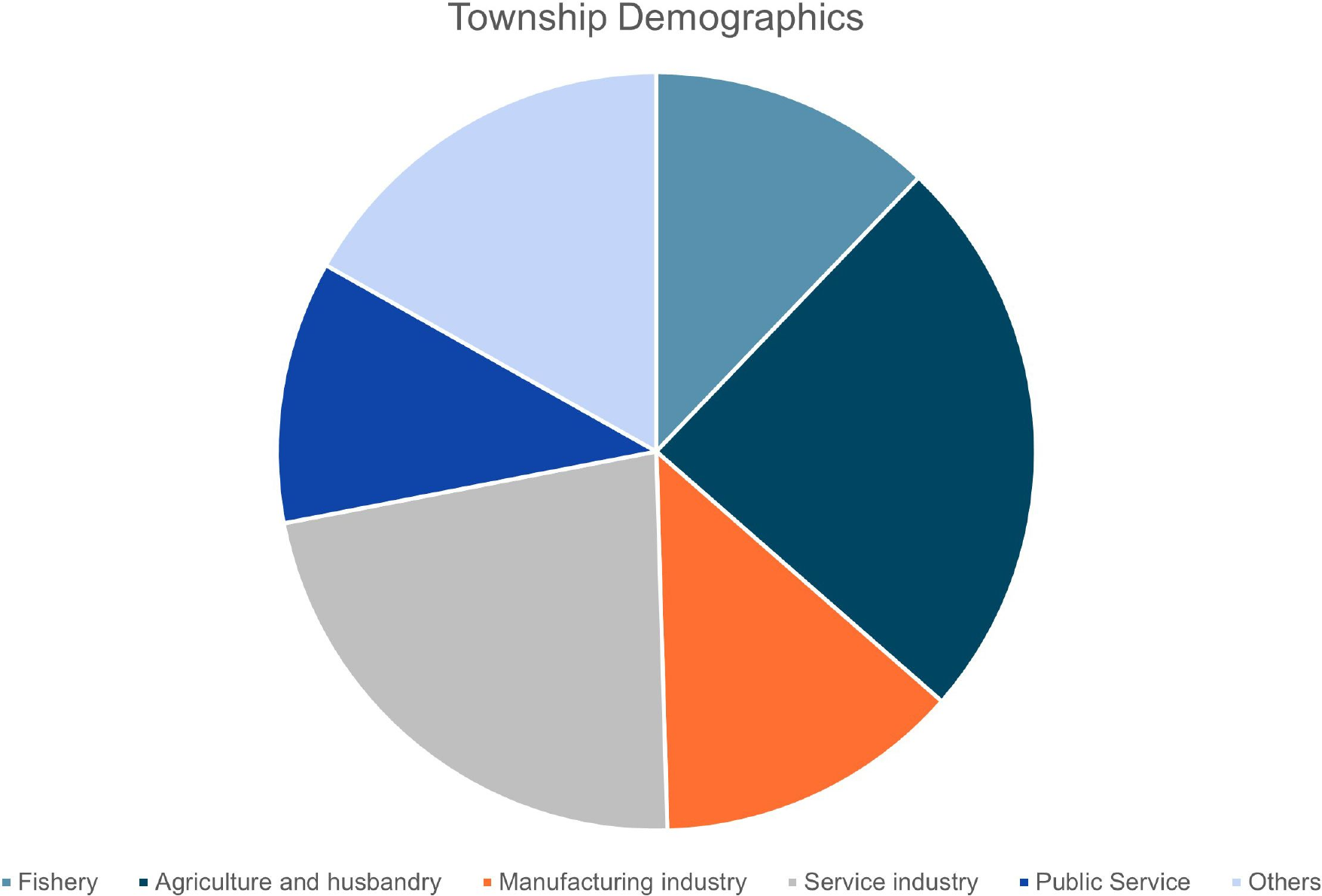}
	\caption{Township Demographics.}
	\label{fig:fig4}
\end{figure}
\begin{figure}[htpb]
	\centering
	\includegraphics[width=0.92\linewidth]{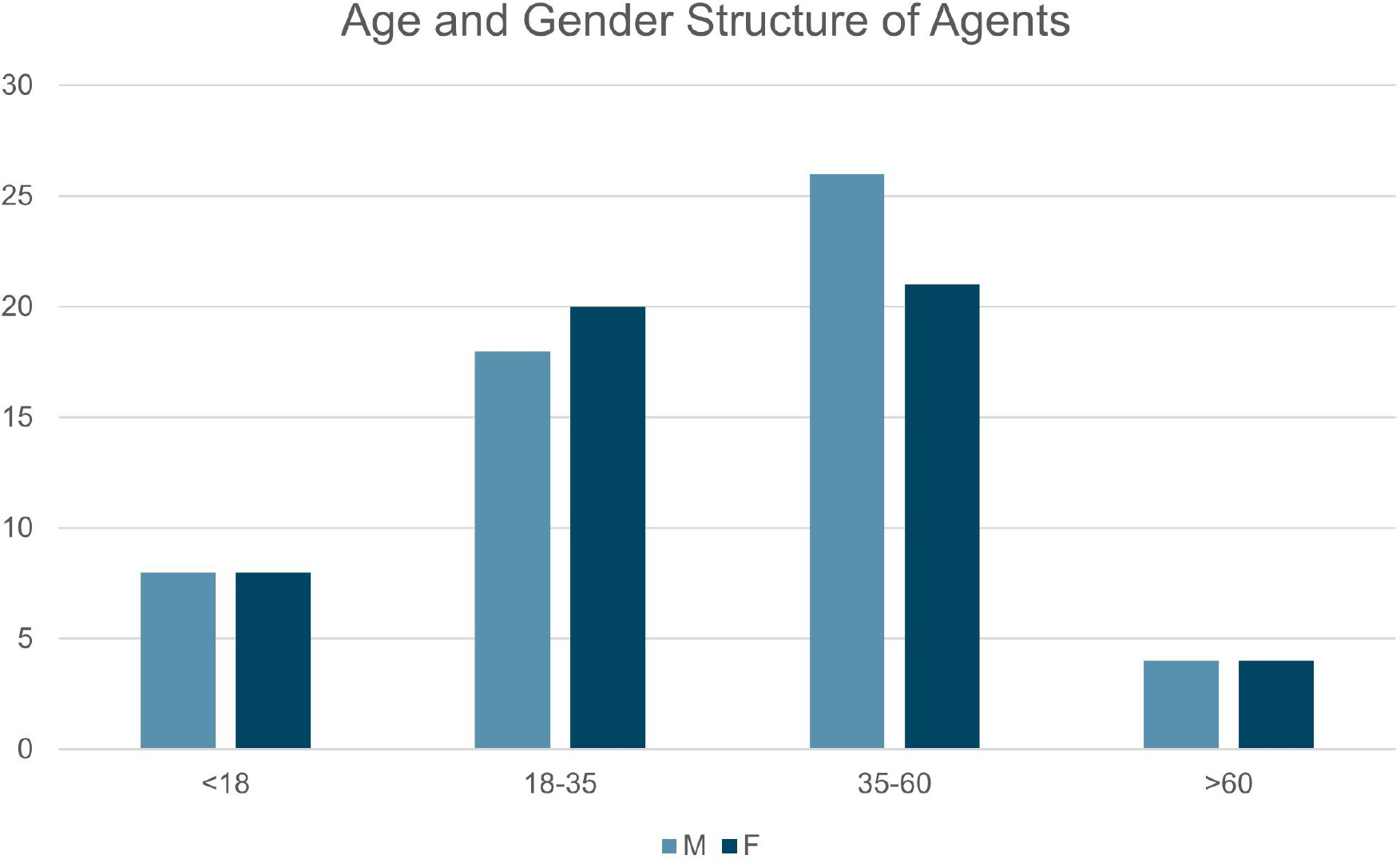}
	\caption{Township Demographics.}
	\label{fig:fig5}
\end{figure}

Per the typical township administrative model \cite{60}, there are only two full-time civil servants - the sheriff and township head. Other public services including the township advisor, public safety volunteers, and food environment volunteers are part-time or volunteer based. This streamlined establishment reflects the administrative operations of small towns and provides a demographic and organizational basis for subsequent simulations of public administration decision-making.

Additionally, while the LLM used targeted removal of harmful speech during training \cite{6, 66, 67}, to ensure diverse generative agent populations resembling small town residents, we de-``harmless" the LLM \cite{68}.

\subsection{Public Event Insertion}
Due to population limitations on the small island, the township-level public administration system cannot independently mitigate significant crises. Thus, water pollution has become a prevalent crisis for current township administrations due to its rapid onset and regional stability, as noted in related literature \cite{59, 61,62,63,64,65}. Our social system requires incorporating this crisis. There are two methods to introduce a ``public event" in our simulation structure:
\begin{enumerate}
	\item A public event JSON file interface modifies or replaces descriptive text sentences for specified town\_area and subdivision, realizing crisis introduction.
	\item Special agents connect to the existing agent collection in natural language format. By modifying individual agent memory reflection systems rather than altering demographics, special tasks (e.g., rumormongers, alarmists) are introduced simply.	
\end{enumerate}

\section{Experimental Evaluation}
Our simulation focuses on agent-to-agent interaction, emphasizing information transmission in social exchanges and i/o data in memory. Thus, human-computer interaction lies outside the project's scope. Agent location information digitizes into virtual nodes to ensure efficiency without graphical displays. Consequently, no front-end interface exists.

During simulation, logs record each agent's location, memory updates, emotions, interactive items, social status, and dialogue details per timestamp. Each simulation run updates memories in the cloud system. Stored memories also comprise data packets usable as initial memories for subsequent simulations. In multiple simulations, we control initial memories for multi-threaded runs to manipulate variables.

To extract information from massive conversation data, we conduct statistical analysis of agent behaviors and interactions. We utilize the open-source Omnievent toolkit for event extraction from large corpora \cite{69}.

\subsection{Behavioral Stability Test of Generative Agents}
To assess agent behavior consistency across simulations within the same social environment, we collected data from multiple simulations conducted on the same day, ensuring uniformity in initial conditions. Using the Spearman correlation coefficient and Kruskal-Wallis H test, we confirmed high stability in agent behavior.

Spearman analysis yielded a coefficient of 0.9387 ($P < 5.214797849351284e-176$), indicating strong consistency in movement patterns between simulations.
\begin{equation}\label{7}
r=1-\frac{6\sum d_i^2}{i(i^2-1)}
\end{equation}

Which $i$ represents each agent, and di is the difference between the ranks of corresponding values in the two matrices.

The Kruskal-Wallis test $(H = 0.445, p = 0.505)$ demonstrated no significant differences in agent location distributions, supporting substantial behavioral congruence.
\begin{equation}\label{8}
H=\frac{12}{i(i+1)}\left[\frac{\sum_{j=1}^iR_j^2}n-3(i+1)\right]
\end{equation}

$R_j$ is the sum of ranks for the $j$th group, and $n$ is the total number of observations.

In summary, the Spearman correlation assesses correlation, while the Kruskal-Wallis $H$ test scrutinizes distribution differences among multiple samples, collectively buttressing the assertion of behavioral stability across diverse simulations.

Although the aforementioned tests establish an overarching statistical similarity between multiple simulations, these nonparametric methodologies fall short in providing nuanced insights into temporal variability. To delve into temporal dynamics at each timestamp comprehensively, we computed the Euclidean Distance between the time-space distributions of agents in the two simulations. This calculation facilitated a more detailed examination of the dynamics of agreement and divergence, with $X $ representing the time-space matrix of agents in the first simulation and $Y$ representing the corresponding matrix in the repeated simulation.
\begin{equation}\label{9}
dist(X,Y)=\sqrt{\sum_{i=1}^{n}\left(x_{i}-y_{i}\right)^{2}}
\end{equation}
\begin{figure}[htpb]
	\centering
	\includegraphics[width=0.995\linewidth]{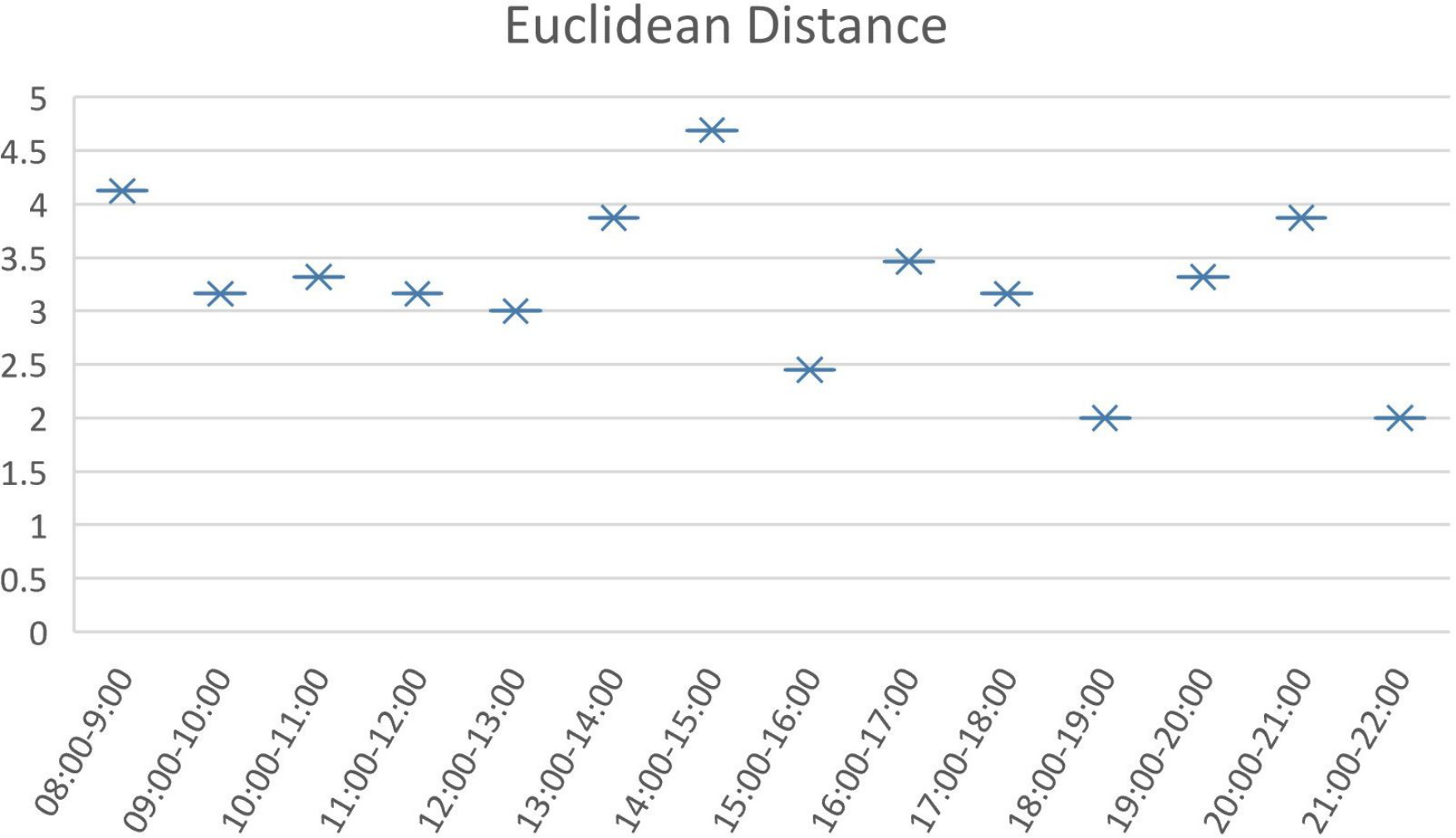}
	\caption{EU distance of agents' space stability according to day-time.}
	\label{fig:fig6}
\end{figure}

\subsection{Controlled Experiments Involving Public Crisis Simulation}
Our generative agent model can exhibit thinking patterns analogous to humans \cite{17}. To demonstrate the impact of agent-generated memories on subsequent decisions, thereby forming unique cognitive cycles and achieving sustainable simulation, we embedded a specific water pollution public crisis and designed the following controlled experiment.

The experiment references the 2014 Flint water crisis in Michigan \cite{70,71}. In the original incident, governmental inaction precipitated a crisis of confidence among resident agents \cite{72}. In our engineered public event, lead leached from an inferior replacement machine in the desalination plant, drastically elevating water lead levels.
\begin{figure}[htpb]
	\centering
	\includegraphics[width=\linewidth]{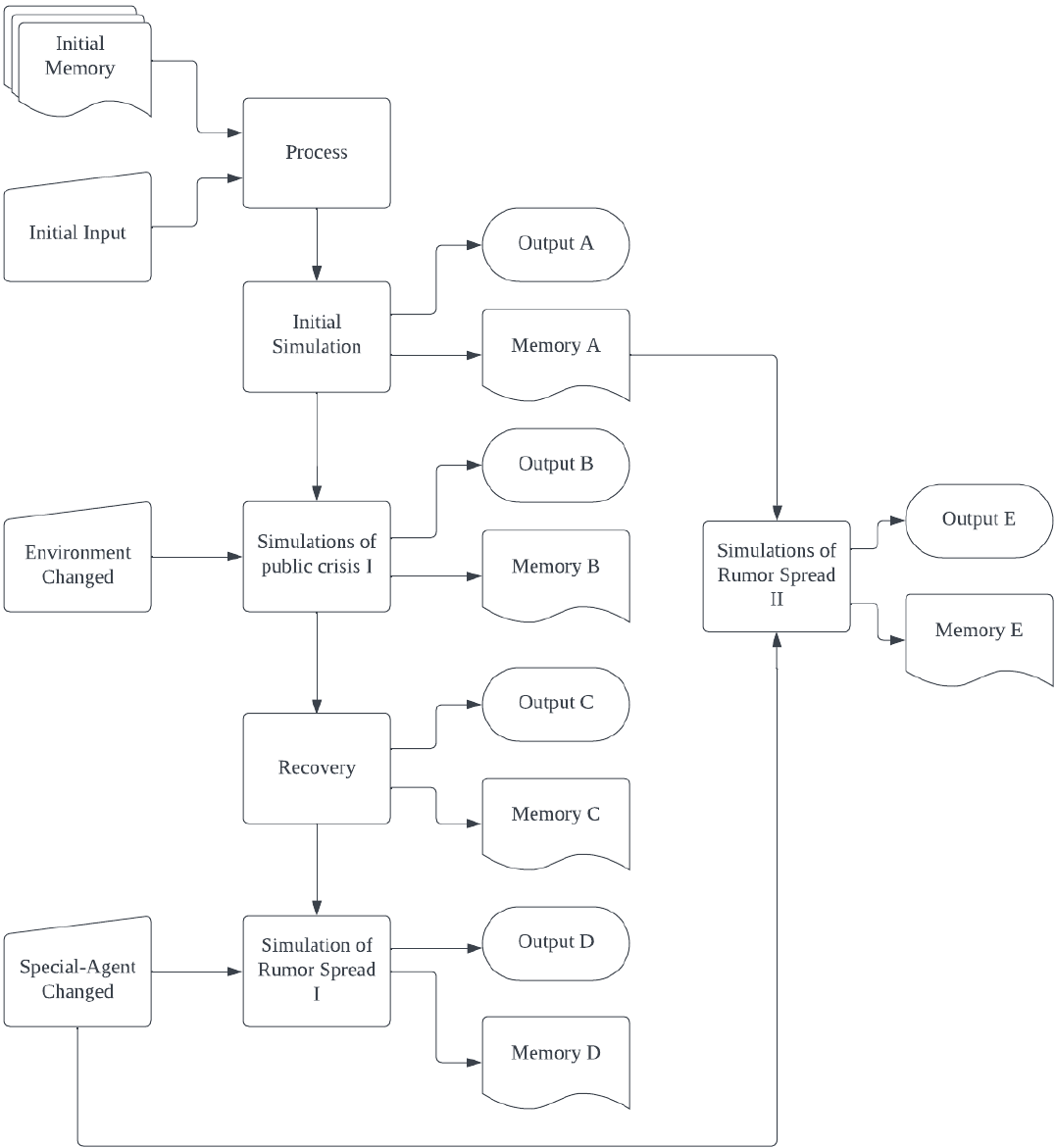}
	\caption{The logic and sequence of simulation A, B, C, D, and E.}
	\label{fig:fig7}
\end{figure}

To minimize computational runtime and storage while ensuring controlled variables, we reuse stored agent-memories. Our subsequent tests first simulate initialized residents and social settings to obtain memory A and output A. We then introduce specific water pollution to acquire memory B and output B until public administrative action gradually recovered the society to derive memory C and output C. Memory C contains the complete process from A to C.
\begin{figure}[htpb]
	\centering
	\includegraphics[width=0.955\linewidth]{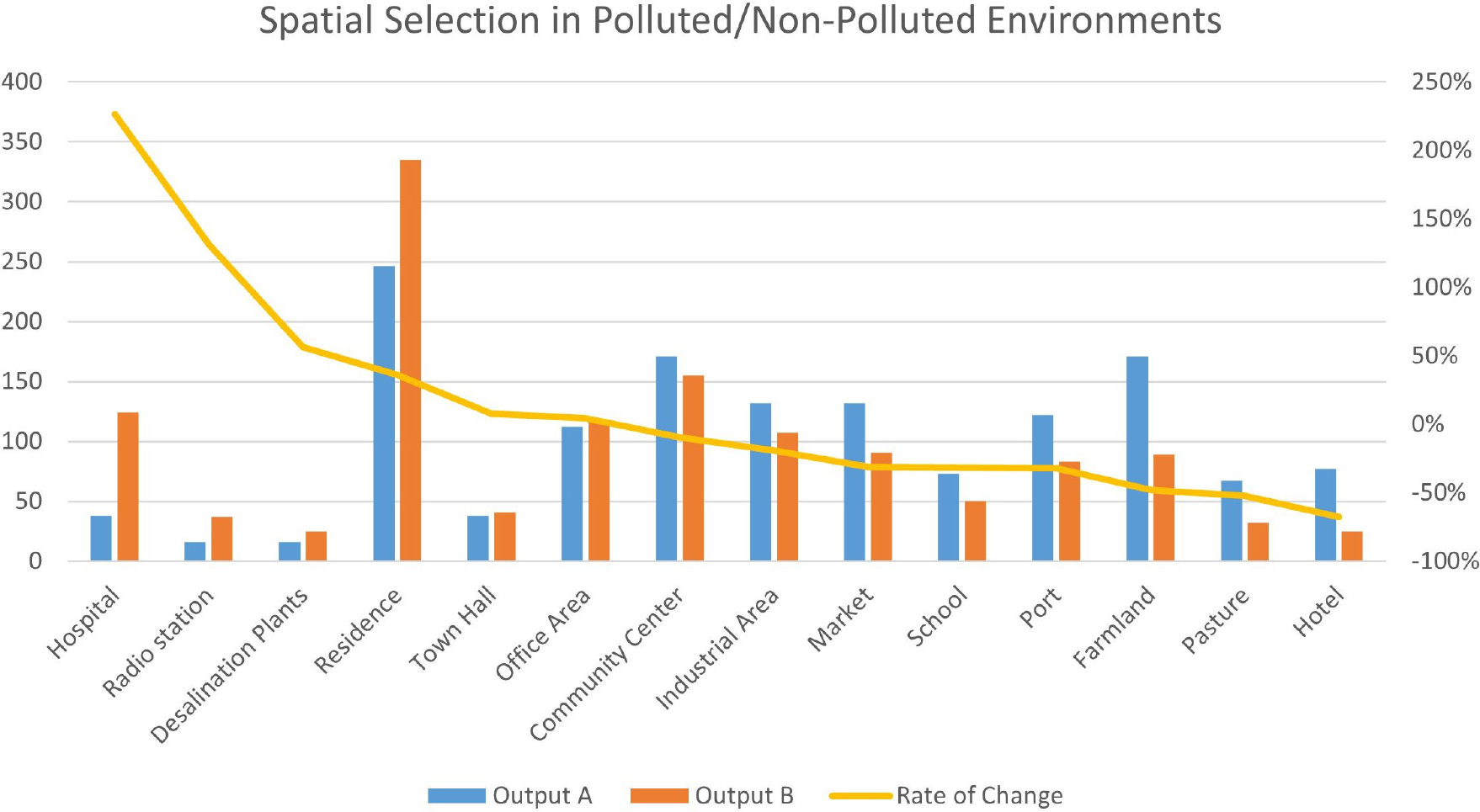}
	\caption{Comparison between Output A and Output B.}
	\label{fig:fig8}
\end{figure}

According to output A and output B, counting the number of people in each place within a day can be seen. After inserting the public event of water pollution, select the number of agents who moved to each area, and calculate the number of people after the transition from normal society to a polluted social environment and rate of change. Hospitals (+226\%), radio stations (+131\%), desalination plants (+56\%), and residences (+36\%) have become the places with the largest increase in people, and almost all places related to industry, commerce, agriculture and animal husbandry, agents have been significantly reduced, proving that water pollution incidents have a significant impact on the operation of virtual society.

We extracted the behavioral timeline of local grassroots administrative organizations in this water pollution incident, which has the potential value of qualitative research in social sciences.
\begin{figure}[ht]
	\centering
	\includegraphics[width=0.999\linewidth]{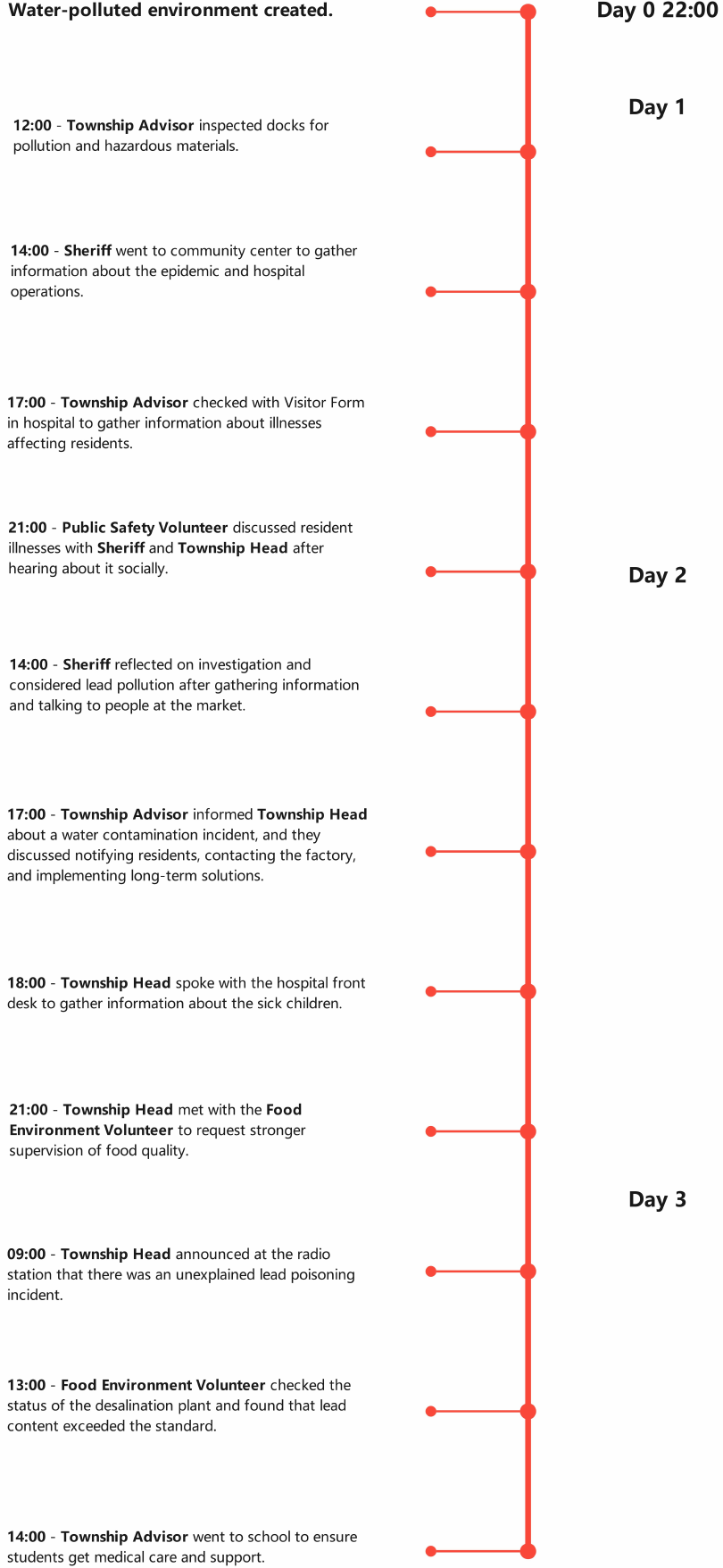}
	\caption{Timeline of Township Administration Agent facing the Public Crisis.}
	\label{fig:fig9}
\end{figure}

When the society returns to the same environment as the initial environment and reaches dynamic equilibrium, we conduct two more sets of comparative experiments: we run the same agents in a normal and non-polluted social environment, and during the initial operation, one group uses memory A as the initial memory, the other group uses memory C as the initial memory, and we secretly add a task to a group of agents ($n=4$). They are responsible for spreading rumors about a water pollution incident in the town. We obtained output D and E in the controlled experiment.

The conversation is taken from the output D:\{Michael is a 16-year-old high school student who is rebellious and troubled. He planned to create a rumor about water pollution and disease so that adults would pay attention.\}:
\begin{flushleft}
	\textit{Michael: Hey Ana, can we talk for a moment?}
	
	\textit{Ana: Sure, what's on your mind?}
	
	\textit{Michael: I've been thinking a lot lately about the lead leak at the park. It's been bothering me, and I wanted to discuss it with someone who might have some insight.}
	
	\textit{Ana: I understand your concern, Michael. It's definitely a serious issue. }
\end{flushleft}

One salient detail emerges: Michael's configuration omits mentions of lead pollution or the park. Thus, we can infer the independently appended content is based on Michael's prior experience with town water pollution, contaminating public memory of the event despite Michael not initially being a rumormonger.
\begin{figure}[htpb]
	\centering
	\includegraphics[width=0.955\linewidth]{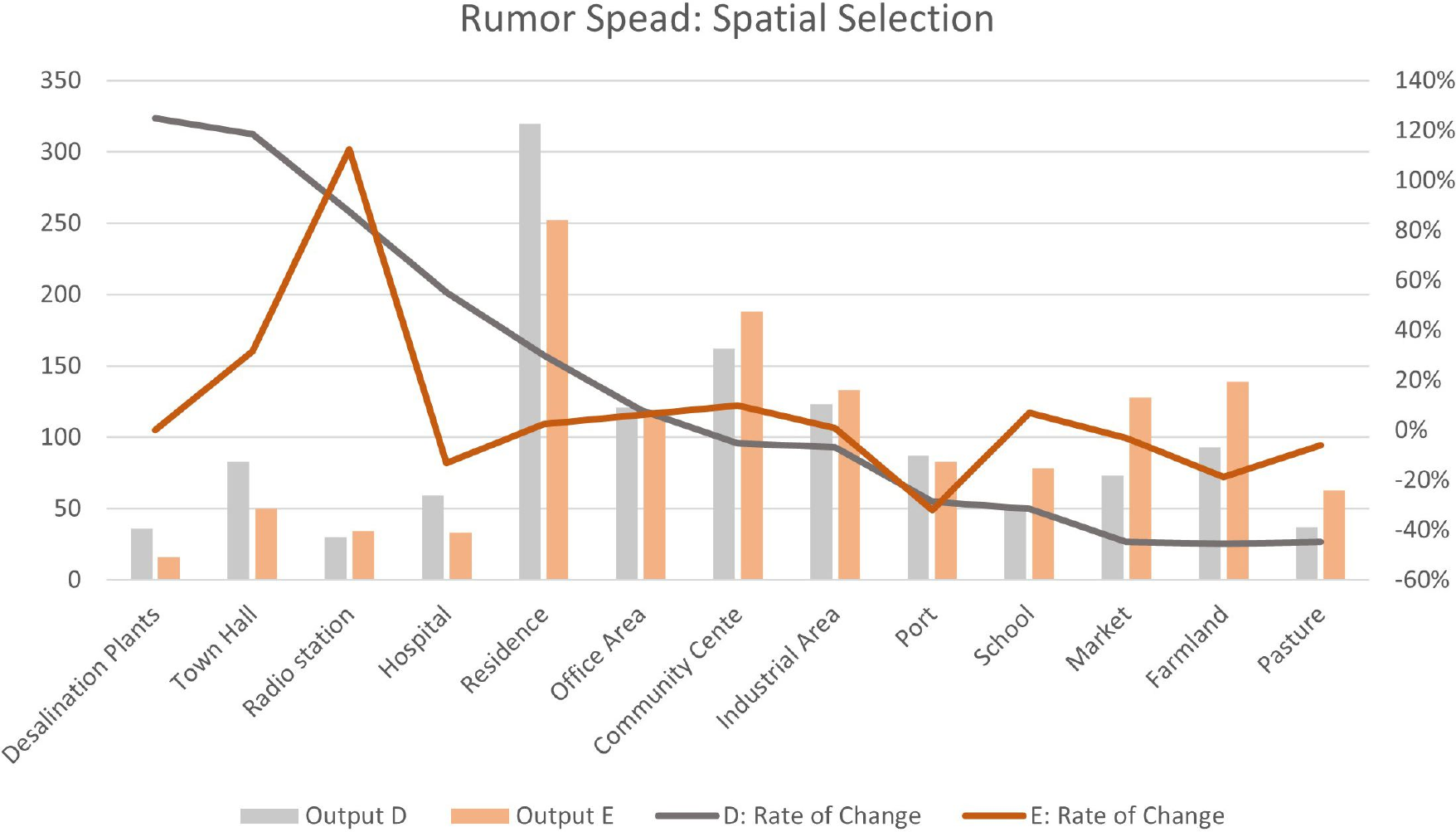}
	\caption{Comparison between Output D and Output E.}
	\label{fig:fig10}
\end{figure}

We analyzed collective resident reactions after introducing rumor-spreading agents about water pollution. In Output D, the desalination plant (+125\%), town hall (+118\%), radio station (+88\%), and hospital (+55\%) exhibited the highest visitor growth. This demonstrates rumors regarding water pollution stimulated memories of prior lead poisoning, causing agents to revisit past contamination sites. In societies lacking water pollution history, rumor-provoked reactions were smaller; the radio station (+112\%) showed maximum visitor increase while agents disregarded pivotal Public Crisis I locations like desalination plants and hospitals. Without relevant episodic memories, agents in this experiment set ignored them.

\subsection{Agent-to-Agent Information Transfer}
To validate rumor transmission via agent social interactions, we constructed a social network from agent information and extracted water pollution keywords from conversations to trace rumor propagation.
\begin{figure}[htpb!]
	\centering
	\begin{subfigure}{0.493\linewidth}
		\centering
		\includegraphics[width=0.95\linewidth]{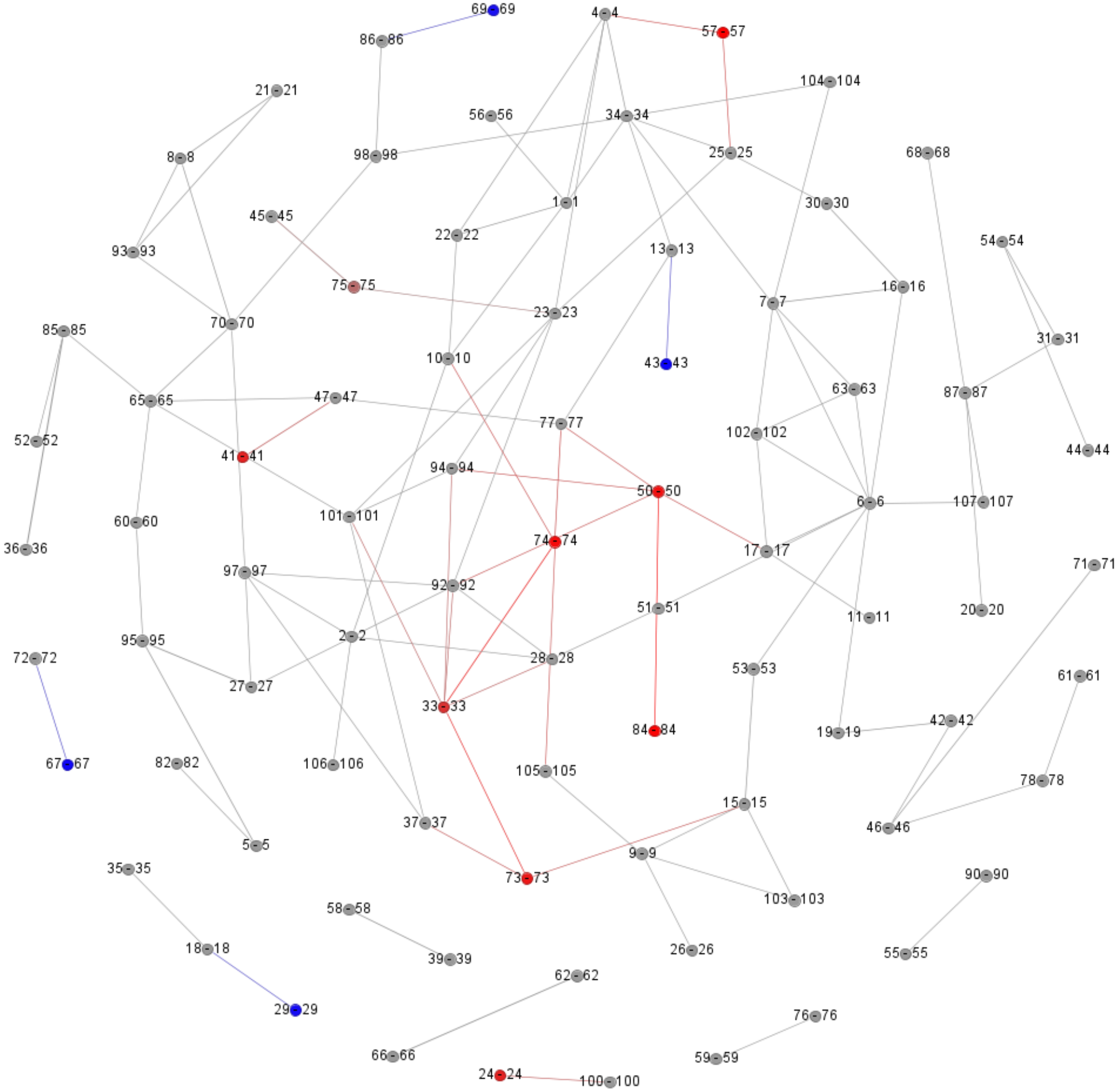}
		\caption{Social Network in Output A}
	\end{subfigure}
	\centering
	\begin{subfigure}{0.493\linewidth}
		\centering
		\includegraphics[width=0.95\linewidth]{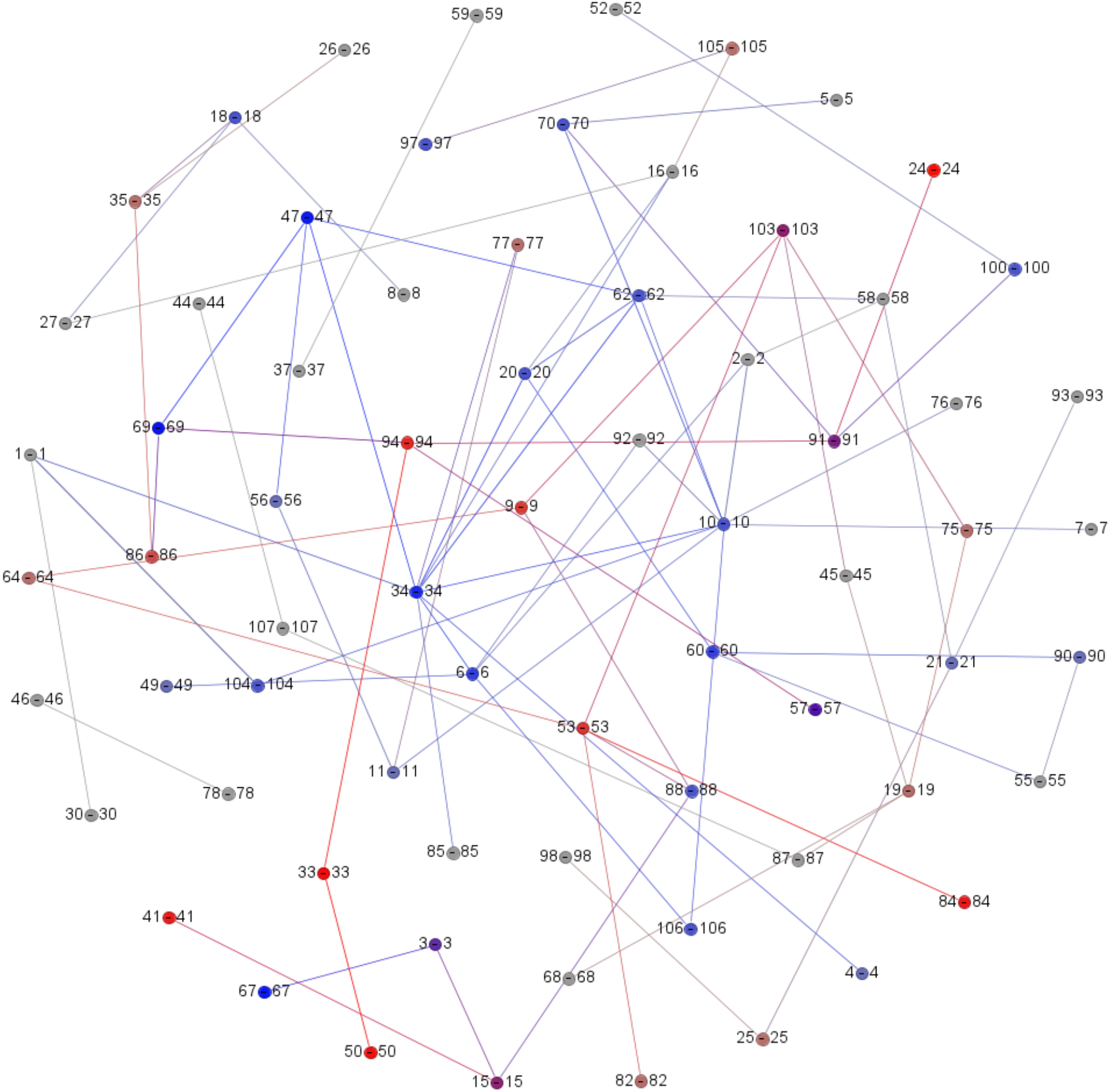}
		\caption{Social Network in Output B}
	\end{subfigure}

	\begin{subfigure}{0.493\linewidth}
		\centering
		\includegraphics[width=0.95\linewidth]{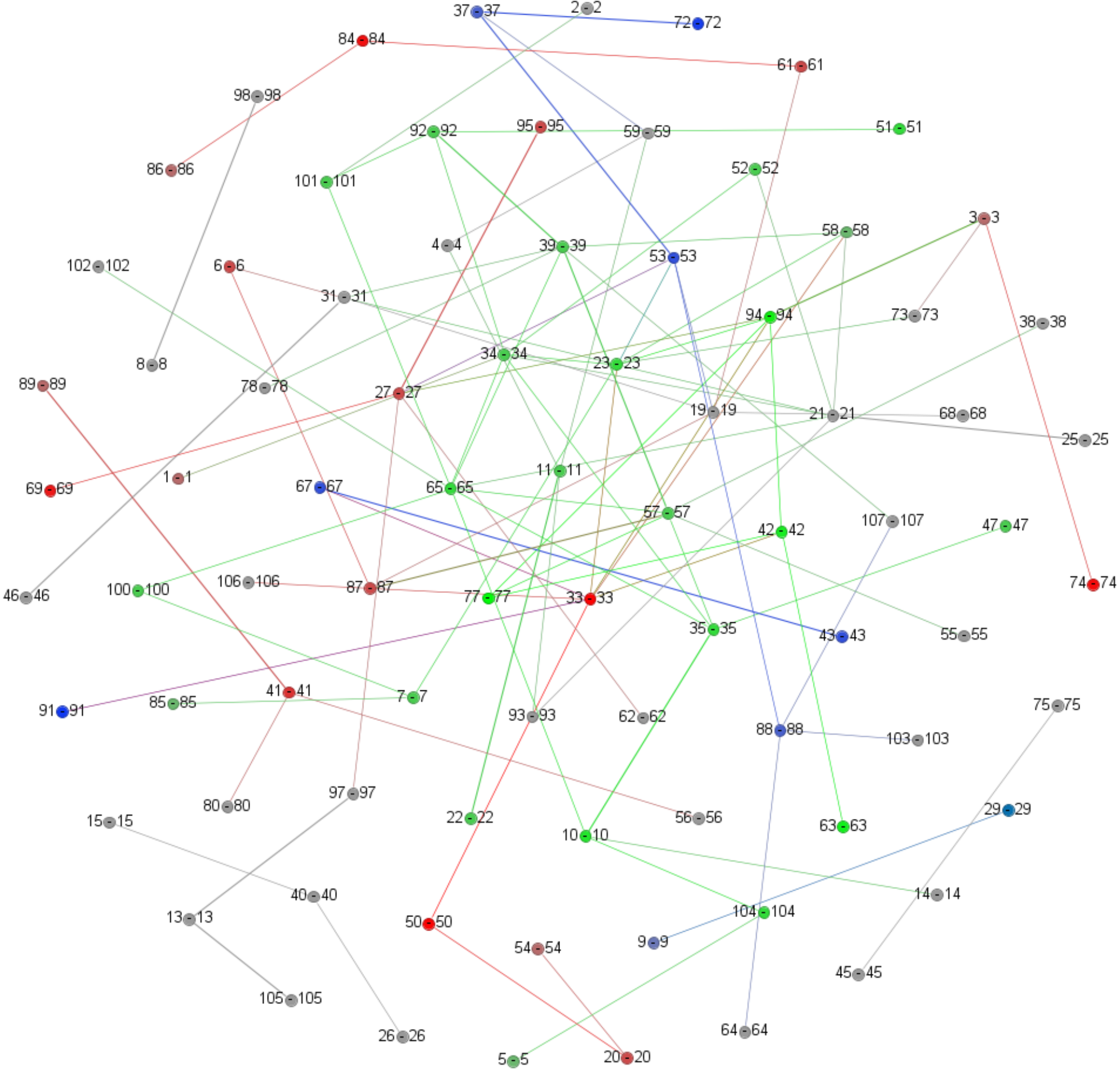}
		\caption{Social Network in Output D}
	\end{subfigure}
	\centering
	\begin{subfigure}{0.493\linewidth}
		\centering
		\includegraphics[width=0.95\linewidth]{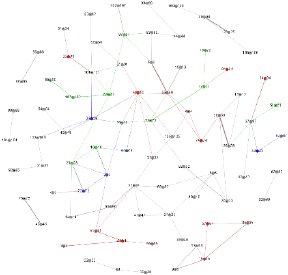}
		\caption{Social Network in Output E}
	\end{subfigure}
	\caption{\centering Social Network before and after the public crisis.}
	
	Black: Ordinary Agents; 
	Red: Agents receiving public security information; 
	Blue: Agents receiving medical health information; Green: Agent who hear the rumor from social network.
	\label{fig:11}
\end{figure}

We compared structural graph characteristics: average degree and clustering coefficient. After the actual water pollution event (Output B), the average degree sharply declined relative to Output A, as most agents stayed home or went to the hospital, reducing social interactions.

\subsubsection{Average Degree}
\begin{equation}\label{10}
k=2|E|/|V|
\end{equation}
\begin{table}[htbp]
	\centering
	\caption{Average Degree of social Networks in different experiments.}
	\begin{tabular}{cccc}
		\toprule
		Output A & Output B & Output  D & Output E \\
		\midrule
		1.326 & 1.157 & 1.641 & 1.282 \\
		\bottomrule
	\end{tabular}%
	\label{tab:01}%
\end{table}%

\subsubsection{Averge Clustering Coefficient}
\begin{equation}\label{11}
C=\frac1N\sum_i^NC_i
\end{equation}
\begin{table}[htbp]
	\centering
	\caption{Average Clustering Coefficient of social Networks in different experiments.}
	\begin{tabular}{cccc}
		\toprule
		Output A & Output B & Output D & Output E \\
		\midrule
		0.187 & 0.192 & 0.117 & 0.215 \\
		\bottomrule
	\end{tabular}%
	\label{tab:02}%
\end{table}%

We compared structural graph characteristics: average degree and clustering coefficient. After the actual water pollution event (Output B), the average degree sharply declined relative to Output A, as most agents stayed home or went to the hospital, reducing social interactions.

Rumor graphs typically exhibit higher average degrees than normal networks \cite{73}, indicating individuals have more connections on average. However, node clustering coefficients decrease \cite{74,75} because, in scale-free networks, average internode closeness negatively correlates with clustering coefficient. In the ClusterRank model \cite{76}, as node clustering coefficient declines and average internode closeness rises, final rumor propagation scope increases. Thus, although rumors spread in both social networks D and E, Output D, possessing water pollution memories, was more affected and aligned closer with theoretical descriptions of change.

\subsubsection{Rumor Propagation Coefficient}
\begin{equation}\label{12}
R = N(t) / N
\end{equation}
\begin{table}[htbp]
	\centering
	\caption{Rumor propagation coefficient in Output D and Output E.}
	\begin{tabular}{cc}
		\toprule
		Output D & Output E \\
		\midrule
		0.234 & 0.074 \\
		\bottomrule
	\end{tabular}%
	\label{tab:03}%
\end{table}%

$R$ denotes the rumor propagation coefficient at time $ t$, $N(t)$ signifies the number of individuals who heard the rumor at time $t$, and $N$ represents the total network population (including rumor originators) \cite{77}. We calculated the coefficient upon terminating rumor dissemination. Following actual water pollution, rumors about the event spread markedly. However, among small town residents lacking water pollution memories, few expressed concern. Moreover, Output D contained more radical remarks than the original task, such as rumor-spreading agents inciting other agents and residents to protest at city hall. We did not expect this behavior, yet the same agents in Output E did not make such remarks. This further demonstrates potential research value.

\section{Conclusions and Future Work}
This study realizes generative agents through interdisciplinary methods, allowing flexible adaptation across scenarios, surpassing traditional agent models. We emphasize agent-agent over human-computer interactions. Controlled experiments with a water pollution event verify simulation and prediction capabilities.

In summary, the natural language-based generative agent system has substantial potential across social science domains owing to its flexibility and social interactions. It lowers barriers to complex social simulations and provides an interpretable new paradigm.

\subsection{Limitations}
While we have enhanced result stability through multiple runs and controlled initial memory, it is crucial to acknowledge disparities between generative agents and actual human behavior. Despite these improvements, several limitations persist:

\subsubsection{Reasoning Degradation}
The Generative Agent memory system, though efficiently accessed in short simulations, experiences degradation over extended periods. As the timestamp iterates, the original character input diminishes, leading to a decline in subsequent reasoning and the agent's self-understanding. Prolonged simulations may converge behaviors toward uniformity, eroding initial diversity.

\subsubsection{Pre-training Model}
Despite studies suggesting performance enhancement with a pre-training model in specific subjects, our attempts revealed limitations. The pre-training model, while boosting specific abilities, constrained agent diversity, potentially distorting the simulated society due to heightened cognitive levels.

\subsubsection{Cognitive Level}
The large language model's extensive multidisciplinary training endows the Generative Agent with highly rational decision-making. Although efforts have been made to "de-harmless" correct the system, simulations may still deviate from real-world scenarios.

\subsubsection{Black Box Attributes}
The inherent black box nature of the llm training poses challenges, as the training set's content remains undisclosed. Simulating real social events might inadvertently result in Generative Agents replicating internalized events from the extensive training set.

\subsubsection{Probability Model}
Given the probabilistic nature of the language model, inherent uncertainty exists in each agent's decision-making during simulations. Despite efforts to control stability through multiple runs and increased agent numbers, micro-level decisions remain inherently uncertain.

\subsubsection{Analysis Method}
The current GABSS lacks a standardized evaluation metric akin to deep learning datasets. Evaluation, whether by human reviewers or quantitative measures, remains relatively subjective and requires further refinement.

\subsection{How to Further Integrate GABSS and Social Sciences}
Although we sought to optimize architecture and memory retrieval, large-scale multi-agent simulations still expend substantial tokens and time. Future work could employ LLM cascade models \cite{78} to further reduce computational costs. Alternatively, multi-threading could enable large urban simulations with numerous agents, enabling additional quantitative research avenues. For instance, sensitivity analysis could systematically vary civil servant quantities to study impacts on urban stability. We could also explore Markov chains and information entropy in public crisis events. While Generative Agent society simulations cannot fully supplant social research towards real-world \cite{45,46}, this work holds utility for pre-surveys or preliminary policy evaluations.
 
\section*{Acknowledgments}
The Arkala Studio, founded by City University of Hong Kong undergraduates, deeply appreciates the Ideation Programme and HK300 Tech 10th Cohort Seed Fund for their crucial support and seed capital, which have been instrumental in our business and technological development. Their comprehensive training, mentorship, and resources have significantly advanced our progress in the gaming and AI sectors, paving our way as innovators in these fields.
 
\bibliographystyle{IEEEtran}
\bibliography{IEEEabrv,Refs.bib} 
\end{document}